\documentclass[11pt]{article}

\title{Graded Lie Algebra Generating of Parastatistical Algebraic Structure
\thanks{This project supported by National Natural Science Foundation of
China and LWTZ 1298.}}
\author{{ Weimin Yang and Sicong Jing}\\
{\small Department of Modern Physics, University of Science and}\\
{\small Technology of China, Hefei 230026, P.R.China}}

\begin{document}%
 
\maketitle


\begin{abstract}
A new kind of graded Lie algebra (we call it $Z_{2,2}$ graded Lie algebra) is
introduced as a framework for formulating parasupersymmetric theories. By
choosing suitable bose subspace of the $Z_{2,2}$ graded Lie algebra and using
relevant generalized Jacobi identities, we generate the whole algebraic
structure of parastatistics.
\end{abstract}

\section{Introduction}
Generalized statistics and supersymmetry are two main interests of theoretical
physics in recent years. Generalized statistics first was introduced in the
form of parastatistics as an exotic possibility extending the Bose and Fermi
statistics \cite{s1}. A basic idea of supersymmetry is the mixture of
particles of different statistics, normally taken to be bosons and fermions
\cite{s2}. Though supersymmetry and parastatistics may be unified in the form
of parasupersymmetry \cite{s3}, nevertheless, by the algebraic construction,
the two concepts seem to be independent.
\par
It is well known that mathematical basis of supersymmetry is $Z_2$ graded Lie
algebra. In this kind of Lie algebra, although not only commutators but also
anticommutators are involved, the basic algebraic structure is still bilinear.
The characteristic of algebraic relations of parastatistics, however, is
trilinear commutation relations, or double commutators and anticommutators.
This will cause some intrinsic difficulties to formulating the
parasupersymmetric theories within a framework of the usual $Z_2$ graded Lie
algebra.
\par
In order to provide a suitable framework to describe parasupersymmetric
theories, we introduce a new kind of graded Lie algebra (maybe call it
$Z_{2,2}$ graded Lie algebra) in this letter, which can be considered as a
generalization or extension of the ordinary $Z_2$ graded Lie algebra. In
contrast with the latter in which there are only two subspaces (bose and fermi
or even and odd subspaces), the former has four subspaces which can be called
bose, fermi, parabose and parafermi subspaces respectively (see the last
section for the reason). We would like to point out that by choosing
appropriate bose subspace of the $Z_{2,2}$ graded Lie algebra and using
relevant Jacobi identities, one can derive all the algebraic relations of a
system consisting of parabosons and parafermions. Therefore, one may analyse
supersymmetric properties of such a parasystem more effectively and more
conveniently on the basis of $Z_{2,2}$ graded Lie algebra.
\par
This letter is arranged as follows. In section 2 we introduce a formal
definition of the $Z_{2,2}$ graded Lie algebra and discuss main differences
between $Z_2$ and $Z_{2,2}$ graded Lie algebras. As an example, in section 3
we choose an algebra $u(1,1)$ as the bose subspace of $Z_{2,2}$ and construct
a whole $Z_{2,2}$ grading of $u(1,1)$ by virtue of the generalized Jacobi
identities. Then in section 4 we demonstrate the derived $Z_{2,2}$ graded Lie
algebra include all the algebraic relations of parastatistics for a system
with one parabose and one parafermi degree of freedom. Some remarks are also
in the last section.

\section{$Z_{2,2}$ graded Lie algebra}
\par
Let ${\bf L}$ be a vector space over a field ${\bf K}$, which is a direct sum
of four subspaces $L_{ij}$ (i,j=0,1), i.e.,
\begin{equation}
{\bf L} = L_{00} \oplus L_{01} \oplus L_{10} \oplus L_{11}.
\end{equation}
For any two generators (vectors) in ${\bf L}$, we define a composition (or
product) rule, written $\circ$, with the following properties
\par
(i) Closure: For $\forall u,v \in {\bf L}$, we have $u \circ v \in {\bf L}$,
i.e.,
\begin{equation}
{\bf L} \times {\bf L} \rightarrow {\bf L}.
\end{equation}
\par
(ii) Bilinearity: For $\forall u,v,w \in {\bf L}$, $c_1,c_2 \in {\bf K}$, we
have
\begin{eqnarray}
(c_1 u + c_2 v) \circ w = c_1 u \circ w + c_2 v \circ w,  \nonumber\\
w \circ (c_1 u + c_2 v) = c_1 w \circ u + c_2 w \circ v.
\end{eqnarray}
\par
(iii) Grading: For $\forall u \in L_{ij}, v \in L_{mn}, (i,j,m,n = 0,1)$, we
have
\begin{equation}
u \circ v = w \in L_{(i+m) mod 2, (j+n) mod 2}.
\end{equation}
For instance, $L_{00} \times L_{00} \rightarrow L_{00}$, $L_{01} \times L_{10}
\rightarrow L_{11}$, $L_{10} \times L_{11} \rightarrow L_{01}$, $\cdots$.
\par
(iv) Supersymmetrization: For $\forall u \in L_{ij}, v \in L_{mn}$, we have
\begin{equation}
u \circ v = - (-1)^{g(u) \cdot g(v)} v \circ u,
\end{equation}
here we assign to any $u \in L_{ij}$ a degree $g(u) = (i,j)$ which satisfies
\begin{equation}
g(u) \cdot g(v) = (i,j) \cdot (m,n) = im + jn,
\end{equation}
and
\begin{equation}
g(u) + g(v) = (i,j) + (m,n) = (i+m,j+n),
\end{equation}
where $v \in L_{mn}$. Obviously, the $g(u)$ looks like a two-dimensional
vector and the expressions in Eqs.(6)-(7) are exactly dot product and additive
operations of the two-dimensional vectors.
\par
(v) Generalized Jacobi identities: For $\forall u \in L_{ij}, v \in L_{kl},
w \in L_{mn}, (i,j,k,l,m,n =0,1)$, we have
\begin{equation}
u\circ (v \circ w) (-1)^{g(u) \cdot g(w)} + v \circ (w \circ u) (-1)^{g(v)
\cdot g(u)} + w \circ (u \circ v) (-1)^{g(w) \cdot g(v)} = 0.
\end{equation}
It is easily to know that there are totally 20 different possibilities for
constructing the generalized Jacobi identities from 4 subspaces of ${\bf L}$ 
$(C_{4}^{1} + C_{4}^{1} C_{3}^{2} + C_{4}^{3} = 20).$
\par
This completes the definition of $Z_{2,2}$ graded Lie algebra. We now
discuss it in more detail. Firstly, we define the product on ${\bf L}$
$(\circ : {\bf L} \times {\bf L} \rightarrow {\bf L})$ as
\begin{equation}
\circ : u \circ v = uv - (-1)^{g(u) \cdot g(v)} vu,
\end{equation}
for $\forall u,v \in {\bf L}$. One can convince oneself that the composition
law (8) is a product which satisfies all conditions of the product of a
$Z_{2,2}$ graded Lie algebra as defined by Eqs.(2-7). Then we consider this
product separately on the subspaces $L_{00}, L_{01}, L_{10}, L{11}$, and
between them. Let generators $X_{ij} and X_{ij}^{\prime}$ belong to the
subspace $L_{ij}$.According to Eqs.(4) and (9), we can write out the following
ten products for ten different generator combinations
\begin{eqnarray}
&\circ&: L_{00} \times L_{00} \rightarrow L_{00},~~~
X_{00} \circ X_{00}^{\prime} = [X_{00}, X_{00}^{\prime}]; \nonumber\\
&\circ&: L_{01} \times L_{01} \rightarrow L_{00},~~~
X_{01} \circ X_{01}^{\prime} = \{ X_{01}, X_{01}^{\prime} \};\nonumber\\
&\circ&: L_{10} \times L_{10} \rightarrow L_{00},~~~
X_{10} \circ X_{10}^{\prime} = \{ X_{10}, X_{10}^{\prime} \};\nonumber\\
&\circ&: L_{11} \times L_{11} \rightarrow L_{00},~~~
X_{11} \circ X_{11}^{\prime} = [X_{11}, X_{11}^{\prime}]: \nonumber\\
&\circ&: L_{00} \times L_{01} \rightarrow L_{01},~~~
X_{00} \circ X_{01} = [X_{00}, X_{01}]; \nonumber\\
&\circ&: L_{00} \times L_{10} \rightarrow L_{10},~~~
X_{00} \circ X_{10} = [X_{00}, X_{10}]; \nonumber\\
&\circ&: L_{00} \times L_{11} \rightarrow L_{11},~~~
X_{00} \circ X_{11} = [X_{00}, X_{11}]; \nonumber\\
&\circ&: L_{01} \times L_{10} \rightarrow L_{11},~~~
X_{01} \circ X_{10} = [X_{01}, X_{10}]; \nonumber\\
&\circ&: L_{01} \times L_{11} \rightarrow L_{10},~~~
X_{01} \circ X_{11} = \{ X_{01}, X_{11} \}; \nonumber\\
&\circ&: L_{10} \times L_{11} \rightarrow L_{01},~~~
X_{10} \circ X_{11} = \{ X_{10}, X_{11} \}.
\end{eqnarray}
Thus we see that the $Z_{2,2}$ graded Lie algebra indeed generalizes and
extends the notion of $Z_2$ graded Lie algebra in two aspects. One is about
the grading. In $Z_2$ case there are only two subspaces, and in $Z_{2,2}$ case
fouer ones, which require the grading of $Z_{2,2}$ must be some kind of double
grading, or twofold grading. Another is the definition of the degree. In $Z_2$
case the degree is a number (taken value 0 or 1), and in  $Z_{2,2}$ case the
degree is a two-dimensional vector (taken value from the set of (0,0), (0,1),
(1,0) and (1,1)). Here we want to point out that there exists also a
significant difference between these two graded Lie algebras concerning the
generalized Jacobi identities. Mathematically, there are only four
possibilities to generate Jacobi-like identities for three generators $u,v and
w$ in terms of commutators or anticommutators, i.e.,
\begin{eqnarray}
&&[u, [v,w]] +[v, [w,u]] +[w, [u,v]]=0, \nonumber\\
&&[u, \{ v,w \}]+[v, \{w,u \}]+[w, \{u,v \}]=0, \nonumber\\
&&[u, \{v,w \}]+ \{v, [w,u] \}- \{w, [u,v] \}=0, \nonumber\\
&&[u,[v,w]]+ \{v, \{w,u \} \}- \{w, \{u,v \} \}=0.
\end{eqnarray}
Notice that only the first three kinds of identities in Eq.(11) appear in the
generalized Jacobi identities of $Z_2$ graded Lie algebra, however, all the
four kinds of identities can be found in the structure of the generalized
Jacobi identities of $Z_{2,2}$ graded Lie algebra.

\section{$Z_{2,2}$ grading of u(1,1) algebra}
As an example, we discuss the $Z_{2,2}$ grading of u(1,1) algebra, which has
four generators $X_{\mu}, (\mu = 1,2,3,4)$. To do this, we take the u(1,1)
algebra as the subspace $L_{00}$, in which the product is
\begin{equation}
\circ: L_{00} \times L_{00} \rightarrow L_{00}, ~~~[X_{\mu}, X_{\nu}] =
C_{\mu \nu \lambda} X_{\lambda}.
\end{equation}
$C_{\mu \nu \lambda}$ are the structure constants of the u(1,1) algebra, whose
nonzero components are $C_{122} = -C_{212} = 1, C_{133} = -C{313} = -1,
C_{231} = -C_{321} = -2$. We denote generators of $L_{01}, L_{10} and L_{11}$
by $Q_{\alpha} (\alpha = 1,...,dim L_{01}), Y_{i} (i = 1,...,dim L_{10}) and
Z_{j} (j = 1,..., dim L_{11})$, respectively.According to the product rules in
Eq.(10), we may introduce the following coefficient matrices $K_{\mu},
H_{\mu}, s_{\mu}, t_{\mu}, u_{\mu}, v_{\mu}, l_{\alpha}, m_{\alpha} and
n_{\alpha}$, to write explicitly out these products
\begin{eqnarray}
&&\{ Q_{\alpha}, Q_{\beta} \} = (H_{\mu})_{\alpha \beta} X_{\mu},
~~~\{ Y_{i}, Y_{j} \}
= (s_{\mu})_{ij} X_{\mu}, ~~~[Z_{i}, Z_{j}] = (t_{\mu})_{ij} X_{\mu},
\nonumber\\
&&[X_{\mu}, Q_{\alpha}] = (K_{\mu})_{\alpha \beta} Q_{\beta},
~~~[X_{\mu}, Y_{i}] = (u_{\mu})_{ij} Y_{j}, ~~~[X_{\mu}, Z_{i}] =
(v_{\mu})_{ij} Z_{j}, \nonumber\\
&&[Q_{\alpha}, Y_{i}] = (l_{\alpha})_{ij} Z_{j}, ~~~\{ Q_{\alpha}, Z_{i} \} =
(m_{\alpha})_{ij} Y_{j}, ~~~\{ Y_{i}, Z_{j} \} = (n_{\alpha})_{ij} Q_{\alpha}.
\end{eqnarray}
Among these matrices, $H_{\mu} (\mu = 1,2,3,4)$ are symmetric $dim L_{01}
\times dim L_{01}$ matrices, $s_{\mu}$ are symmetric $dim L_{10} \times
dim L_{10}$ ones, and $t_{\mu}$ are antisymmetric $dim L_{11} \times
dim L_{11}$ ones. From the third line of Eq.(13), we observe that the
matrices $l_{\alpha}, m_{\alpha} and n_{\alpha}$ are square matrices only if
$dim L_{10} = dim L_{11}$, so we consider just this case in what follows. Of
course, these matrices are restricted by the generalized Jacobi identities.
Considering all the possible generalized Jacobi identities, we find that these
matrices must satisfy the following constraint relations
\begin{eqnarray}
&& [K_{\mu}, K_{\nu}] = -C_{\mu \nu \lambda} K_{\lambda}, ~~~
[u_{\mu}, u_{\nu}] = -C_{\mu \nu \lambda} u_{\lambda}, ~~~
[v_{\mu}, v_{\nu}] = -C_{\mu \nu \lambda} v_{\lambda}, \nonumber\\
&& K_{\mu} H_{\nu} + (K_{\mu} H_{\nu})^{T} = C_{\mu \lambda \nu} H_{\lambda},
~~~ u_{\mu} s_{\nu} + (u_{\mu} s_{\nu})^{T} = -C_{\mu \lambda \nu}
s_{\lambda},  \nonumber\\
&& v_{\mu} t_{\nu} + t_{\nu} (v_{\mu})^{T} = -C_{\mu \lambda \nu}
(t_{\lambda})^{T}, ~~~
l_{\alpha} v_{\mu} - u_{\mu} l_{\alpha} = (K_{\mu})_{\alpha \beta}
l_{\beta}, \nonumber\\
&& u_{\mu} n_{\alpha} + n_{\alpha} (v_{\mu})^{T} =
(K_{\mu})_{\alpha \beta}^{T} n_{\beta}, ~~~
s_{\mu} (m_{\alpha})^{T} + l_{\alpha} t_{\mu} = (H_{\mu})_{\alpha \beta}
n_{\beta},  \nonumber\\
&& l_{\alpha} (n_{\beta})^{T} + n_{\beta} (l_{\alpha})^{T} =
-(K_{\mu})_{\alpha \beta} s_{\mu}, ~~~
l_{\alpha} m_{\beta} + l_{\beta} m_{\alpha} = (H_{\mu})_{\alpha \beta}
u_{\mu}, \nonumber\\
&& (n_{\alpha})_{ji} (l_{\alpha})_{kl} +
(n_{\alpha})_{ki} (l_{\alpha})_{jl}
= - (s_{\mu})_{jk} (v_{\mu})_{il}, \nonumber\\
&& (n_{\alpha})_{ji} (m_{\alpha})_{kl} - (n_{\alpha})_{jk} (m_{\alpha})_{il}
= (t_{\mu})_{ki} (u_{\mu})_{jl}, \nonumber\\
&& (H_{\mu})_{\alpha \beta} (K_{\mu})){\gamma \delta} +
(H_{\mu})_{\beta \gamma} (K_{\mu})_{\alpha \delta} +
(H_{\mu})_{\gamma \alpha} (K_{\mu})_{\beta \delta} = 0, \nonumber\\
&& (s_{\mu})_{ij} (u_{\mu})_{kl} + (s_{\mu})_{jk} (u_{\mu})_{il}
+ (s_{\mu})_{ki} (u_{\mu})_{jl} = 0, \nonumber\\
&& (t_{\mu})_{ij} (v_{\mu})_{kl} + (t_{\mu})_{jk} (v_{\mu})_{il}
+ (t_{\mu})_{ki} (v_{\mu})_{jl} = 0,
\end{eqnarray}
where the notation "T" means transpose of a matrix, and the ranges of the
indices are $\mu, \nu, \lambda = 1,2,3,4$; $\alpha, \beta, \gamma, \delta =
1,...,dim L_{01}$; $i,j,k,l = 1,...,dim L_{10} (= dim L_{11})$, respectively.
Carefully observing these relations, we realize that $K_{\mu}$ and $H_{\mu}$
have to be $dim L_{01} \times dim L_{01}$ matrices, and all the other ones
are $dim L_{10} \times L_{10}$. For the simplist nontrivial case, i.e.,
$dim L_{01} =4$ and $dim L_{10} = dim L_{11} =2$, by using the trial-and-error
method, we find out the following solutions of these matrices
\begin{eqnarray}
&&K_{1} = \frac{1}{2} \left(
\begin{array}{llll}
-1 & 0 & 0 & 0 \\
 0 & 1 & 0 & 0 \\
 0 & 0 & 1 & 0 \\
 0 & 0 & 0 & -1
\end{array}    \right), ~~~
K_{2} = \left(
\begin{array}{llll}
 0 &  0 & -1 & 0 \\
 0 &  0 &  0 & 0 \\
 0 &  0 &  0 & 0 \\
 0 & -1 &  0 & 0
\end{array}     \right), \nonumber\\
&&K_{3} = \left(
\begin{array}{llll}
 0 & 0 & 0 & 0 \\
 0 & 0 & 0 & 1 \\
 1 & 0 & 0 & 0 \\
 0 & 0 & 0 & 0
\end{array}    \right), ~~~
K_{4} = \frac{1}{2} \left(
\begin{array}{llll}
-1 & 0 &  0 & 0 \\
 0 & 1 &  0 & 0 \\
 0 & 0 & -1 & 0 \\
 0 & 0 &  0 & 1
\end{array}    \right);
\end{eqnarray}
\begin{eqnarray}
&&H_{1} = 2 \left(
\begin{array}{llll}
 0 & 1 & 0 & 0 \\
 1 & 0 & 0 & 0 \\
 0 & 0 & 0 & 1 \\
 0 & 0 & 1 & 0
\end{array}    \right), ~~~
H_{2} = 2 \left(
\begin{array}{llll}
 0 & 0 & 0 & 0 \\
 0 & 0 & 1 & 0 \\
 0 & 1 & 0 & 0 \\
 0 & 0 & 0 & 0
\end{array}    \right), \nonumber\\
&&H_{3} = 2 \left(
\begin{array}{llll}
 0 & 0 & 0 & 1 \\
 0 & 0 & 0 & 0 \\
 0 & 0 & 0 & 0 \\
 1 & 0 & 0 & 0
\end{array}    \right), ~~~
H_{4} = 2 \left(
\begin{array}{llll}
 0 & -1 & 0 & 0 \\
-1 &  0 & 0 & 0 \\
 0 &  0 & 0 & 1 \\
 0 &  0 & 1 & 0
\end{array}    \right);
\end{eqnarray}
\begin{equation}
s_{1} = 4 \left( \begin{array}{ll} 0 & 1 \\ 1 & 0 \end{array} \right), ~
s_{2} = 4 \left( \begin{array}{ll} 1 & 0 \\ 0 & 0 \end{array} \right), ~
s_{3} = 4 \left( \begin{array}{ll} 0 & 0 \\ 0 & 1 \end{array} \right), ~
s_{4} =   \left( \begin{array}{ll} 0 & 0 \\ 0 & 0 \end{array} \right);
\end{equation}
\begin{equation}
t_{1} =   \left( \begin{array}{ll} 0 & 0 \\ 0 & 0 \end{array} \right), ~
t_{2} =   \left( \begin{array}{ll} 0 & 0 \\ 0 & 0 \end{array} \right), ~
t_{3} =   \left( \begin{array}{ll} 0 & 0 \\ 0 & 0 \end{array} \right), ~
t_{4} = 4 \left( \begin{array}{ll} 0 & 1 \\-1 & 0 \end{array} \right);
\end{equation}
\begin{equation}
u_{1} = \frac{1}{2}
          \left( \begin{array}{ll} 1 & 0 \\ 0 &-1 \end{array} \right), ~
u_{2} =   \left( \begin{array}{ll} 0 & 0 \\-1 & 0 \end{array} \right), ~
u_{3} =   \left( \begin{array}{ll} 0 & 1 \\ 0 & 0 \end{array} \right), ~
u_{4} =   \left( \begin{array}{ll} 0 & 0 \\ 0 & 0 \end{array} \right);
\end{equation}
\begin{equation}
v_{1} =   \left( \begin{array}{ll} 0 & 0 \\ 0 & 0 \end{array} \right), ~
v_{2} =   \left( \begin{array}{ll} 0 & 0 \\ 0 & 0 \end{array} \right), ~
v_{3} =   \left( \begin{array}{ll} 0 & 0 \\ 0 & 0 \end{array} \right), ~
v_{4} = \frac{1}{2}
          \left( \begin{array}{ll} 1 & 0 \\ 0 &-1 \end{array} \right);
\end{equation}
\begin{equation}
l_{1} =   \left( \begin{array}{ll} 0 & 1 \\ 0 & 0 \end{array} \right), ~
l_{2} =   \left( \begin{array}{ll} 0 & 0 \\-1 & 0 \end{array} \right), ~
l_{3} =   \left( \begin{array}{ll} 0 & 0 \\ 0 &-1 \end{array} \right), ~
l_{4} =   \left( \begin{array}{ll} 1 & 0 \\ 0 & 0 \end{array} \right);
\end{equation}
\begin{equation}
m_{1} =   \left( \begin{array}{ll} 0 & 1 \\ 0 & 0 \end{array} \right), ~
m_{2} =   \left( \begin{array}{ll} 0 & 0 \\ 1 & 0 \end{array} \right), ~
m_{3} =   \left( \begin{array}{ll} 1 & 0 \\ 0 & 0 \end{array} \right), ~
m_{4} =   \left( \begin{array}{ll} 0 & 0 \\ 0 & 1 \end{array} \right);
\end{equation}
\begin{equation}
n_{1} = 2 \left( \begin{array}{ll} 0 & 0 \\ 0 & 1 \end{array} \right), ~
n_{2} = 2 \left( \begin{array}{ll} 1 & 0 \\ 0 & 0 \end{array} \right), ~
n_{3} = 2 \left( \begin{array}{ll} 0 & 1 \\ 0 & 0 \end{array} \right), ~
n_{4} = 2 \left( \begin{array}{ll} 0 & 0 \\ 1 & 0 \end{array} \right).
\end{equation}
By checking straightforwardly one will believe that these matrices satisfy all
of the constraint relations (14). Thus we obtain the $Z_{2,2}$ grading of
$u(1,1)$ algebra. It has 12 generators: $X_{\mu}$ and $Q_{\alpha}$ for
$\mu, \alpha = 1,2,3,4$, and $Y_{i}$ and $Z_{j}$ for $i, j = 1,2$, which
satisfy the following nonzero commutation and anticommutation relations
\begin{equation}
[X_{1}, X_{2}] = X_{2}, ~~~ [X_{1}, X_{3}] = -X_{3},
~~~ [X_{2}, X_{3}] = -2X_{1};
\end{equation}
\begin{eqnarray}
&& \{ Q_{1}, Q_{2} \} = 2X_{1} -2X_{4}, ~~~\{ Q_{1}, Q_{4} \} = 2X_{3},
\nonumber\\
&& \{ Q_{3}, Q_{4} \} = 2X_{1} +2X_{4}, ~~~\{ Q_{2}, Q_{3} \} = 2X_{2};
\end{eqnarray}
\begin{equation}
\{ Y_{1}, Y_{1} \} = 4X_{2}, ~~~\{ Y_{1}, Y_{2} \} = 4X_{1}, ~~~
\{ Y_{2}, Y_{2} \} = 4X_{3};
\end{equation}
\begin{equation}
[Z_{1}, Z_{2}] = 4X_{4};
\end{equation}
\begin{eqnarray}
&& [X_{1}, Q_{1}] = -\frac{1}{2} Q_{1},
 ~~~ [X_{1}, Q_{2}] = \frac{1}{2} Q_{2},
 ~~~ [X_{1}, Q_{3}] = \frac{1}{2} Q_{3}, \nonumber\\
&& [X_{1}, Q_{4}] = -\frac{1}{2} Q_{4},
 ~~~ [X_{2}, Q_{1}] = -Q_{3},
 ~~~ [X_{2}, Q_{4}] = -Q_{2}, \nonumber\\
&& [X_{3}, Q_{2}] = Q_{4},
 ~~~ [X_{3}, Q_{3}] = Q_{1},
 ~~~ [X_{4}, Q_{1}] = -\frac{1}{2} Q_{1}, \nonumber\\
&& [X_{4}, Q_{2}] = \frac{1}{2} Q_{2},
 ~~~ [X_{4}, Q_{3}] = -\frac{1}{2} Q_{3},
 ~~~ [X_{4}, Q_{4}] = \frac{1}{2} Q_{4};
\end{eqnarray}
\begin{eqnarray}
&& [X_{1}, Y_{1}] = \frac{1}{2} Y_{1}, ~~~ [X_{1}, Y_{2}] = -\frac{1}{2} Y_{2},
\nonumber\\
&& [X_{2}, Y_{2}] = -Y_{1}, ~~~ [X_{3}, Y_{1}] = Y_{2};
\end{eqnarray}
\begin{equation}
[X_{4}, Z_{1}] = \frac{1}{2} Z_{1}, ~~~ [X_{4}, Z_{2}] = -\frac{1}{2} Z_{2};
\end{equation}
\begin{eqnarray}
&& [Q_{1}, Y_{1}] = Z_{2}, ~~~ [Q_{2}, Y_{2}] = -Z_{1}, \nonumber\\
&& [Q_{3}, Y_{2}] = -Z_{2}, ~~~ [Q_{4}, Y_{1}] = Z_{1};
\end{eqnarray}
\begin{eqnarray}
&& \{ Q_{1}, Z_{1} \} = Y_{2}, ~~~\{ Q_{2}, Z_{2} \} = Y_{1}, \nonumber\\
&& \{ Q_{3}, Z_{1} \} = Y_{1}, ~~~\{ Q_{4}, Z_{2} \} = Y_{2};
\end{eqnarray}
\begin{eqnarray}
&& \{ Y_{1}, Z_{1} \} = 2Q_{2}, ~~~\{ Y_{1}, Z_{2} \} = 2Q_{3}, \nonumber\\
&& \{ Y_{2}, Z_{1} \} = 2Q_{4}, ~~~\{ Y_{2}, Z_{2} \} = 2Q_{1}.
\end{eqnarray}

\section{Algebraic structure of parastatistics}
If we identify the generators $Y_{1}$ and $Y_{2}$ in the subspace $L_{10}$
with parabose creation and annihilation operators $a^{\dagger}$ and $a$
respectively, meanwhile, identify the generators $Z_{1}$ and $Z_{2}$ in the
subspace $L_{11}$ with parafermi creation and annihilation operators
$f^{\dagger}$ and $f$ respectively, we will obtain a whole algebraic structure
of a system with one paraboson and one parafermion from the $Z_{2,2}$ graded
Lie algebra of $u(1,1)$ (Eqs.(24)-(33)). In fact, in this case, according to
Eqs.(26) and (27), the four generstors $X_{\mu}$ in the subspace $L_{00}$ can
be written as
\begin{equation}
X_{1} = \frac{1}{4} \{ a^{\dagger}, a \},~ X_{2} = \frac{1}{2} a^{\dagger 2},
~X_{3} = \frac{1}{2} a^{2}, ~X_{4} = \frac{1}{4} [f^{\dagger}, f],
\end{equation}
and the four generators $Q_{\alpha}$ in the $L_{01}$ subspace will have the
forms
\begin{equation}
 Q_{1} = \frac{1}{2} \{a, f \},~
 Q_{2} = \frac{1}{2} \{ a^{\dagger}, f^{\dagger} \},
~Q_{3} = \frac{1}{2} \{ a^{\dagger}, f \},
~Q_{4} = \frac{1}{2} \{ a, f^{\dagger} \},
\end{equation}
from Eq.(33). Obviously, here $X_{1}$ and $X_{4}$ are hermite operators,
$X_{2}$, $Q_{1}$ and $Q_{3}$ are hermitian conjugate to $X_{3}$, $Q_{2}$ and
$Q_{4}$ respectively. Then the  Eq.(29) will give standard trilinear algebraic
relations for paraboson
\begin{equation}
[\{ a^{\dagger}, a \}, a^{\dagger}] = 2a^{\dagger},
~ [\{ a^{\dagger}, a \}, a] = -2a,
~ [a^{\dagger 2}, a] = -2a^{\dagger},~ [a^{2}, a^{\dagger}] = 2a,
\end{equation}
and the Eq.(30) will give standard trilinear algebraic relations for
parafermion
\begin{equation}
[[f^{\dagger}, f], f^{\dagger}] = 2f^{\dagger}, [[f^{\dagger}, f], f] = 2f.
\end{equation}
It is worth pointing out that the Eqs.(31) and (32) will lead to
\begin{eqnarray}
&& [\{ a, f \}, a^{\dagger}] = 2f,~~~ [\{ a^{\dagger}, f \}, a] = -2f,
\nonumber\\
&& \{ \{ a, f \}, f^{\dagger} \} = 2a,~~~
\{ \{a, f^{\dagger} \}, f \} = -2a,
\end{eqnarray}
together with the adjoint ones, which are exactly algebraic relations between
paraboson and parafermion with same parastatistical order $p$ \cite{s4}. The
remaining relations of the $Z_{2,2}$ graded Lie algebra (i.e., Eqs.(24), (25)
and (28)) just mean that the generators $X_{\mu}$ and $Q_{\alpha}$ form a
ordinary $Z_{2}$ graded Lie algebra. Usually the generators $X_{\mu}$ and
$Q_{\alpha}$ are called bose and fermi (or even and odd) generators, and the
subspaces $L_{00}$ and $L_{01}$ the bose and fermi subspaces, respectively.
Therefore, it is reasonable to call the subspaces $L_{10}$ and $L_{11}$ the
parabose and parafermi subspaces respectively.
\par
In summary, we generalize the $Z_{2}$ graded Lie algebra to a more complicated
grading scheme, $Z_{2,2}$ grading, in this letter. By choosing appropriately
the bose subspace of the $Z_{2,2}$ graded Lie algebra, and using the
generalized Jacobi identities, one may derive all parastatistical algebraic
relations. It is worth mentioning that Biswas and Soni also pointed out that
the even and odd generators made up of parabosons and parafermions may give 
operator realizations of graded Lie algebras \cite{s5}. The key difference
between Ref.(5) and this letter is we generalize the ordinary $Z_{2}$ grading
to a new $Z_{2,2}$ grading, so that all of the parastatistical algebraic
relations can automatically appear in our structure. The $Z_{2,2}$ graded Lie
algebra also provides a more potential framework to inverstigate various
possible supersymmetric problem, such as supersymmetry between boson and
parafermion \cite{s6}, bosob and paraboson \cite{s7}, and so on. Work in this
direction is on progrees.

\end{document}